\documentclass[12pt,preprint]{aastex}
\usepackage[]{natbib}
\usepackage{graphics}
\usepackage{emulateapj5}
\usepackage{apjfonts}

\newcommand{\etal}{et al.}  
\newcommand{\per}{\ensuremath{^{-1}}}
\newcommand{\persq}{\ensuremath{^{-2}}}

\newcommand{\msun}{\ensuremath{M_{\odot}}}
\newcommand{\mbh}{\ensuremath{M_\mathrm{BH}}}

\newcommand{\kms}{km s\ensuremath{^{-1}}}

\newcommand{\chisq}{\ensuremath{\chi^2}}
\newcommand{\hst}{\emph{HST}}
\newcommand{\msigma}{\ensuremath{M_{\mathrm{BH}} - \sigma}}
\newcommand{\mgb}{\ion{Mg}{1}\emph{b}}
\newcommand{\hbeta}{H\ensuremath{\beta}}
\newcommand{\ledd}{\ensuremath{L_{\mathrm{Edd}}}}

\newcommand{\hal}{H$\alpha$}
\newcommand{\reff}{\ensuremath{r_e}}
\newcommand{\mue}{\ensuremath{\langle\mu_e\rangle}}

\slugcomment{}
\shorttitle{BLACK HOLES AND HOSTS OF BL LAC OBJECTS} 
\shortauthors{BARTH ET AL.}

\begin{document} 

\title{The Black Hole Masses and Host Galaxies of BL Lac Objects}

\author{Aaron J. Barth\altaffilmark{1,2}, Luis C. Ho\altaffilmark{3},
and Wallace L. W. Sargent\altaffilmark{1}}

\altaffiltext{1}{Palomar Observatory, 105-24 Caltech, Pasadena, CA
91125; barth@astro.caltech.edu}
\altaffiltext{2}{Hubble Fellow}
\altaffiltext{3}{The Observatories of the Carnegie Institution of
Washington, 813 Santa Barbara Street, Pasadena, CA 91101}

\begin{abstract}
 
We have measured the central stellar velocity dispersion in the host
galaxies of 11 BL Lac objects with redshifts $z \leq 0.125$.  The
range of velocity dispersions, $\sim170-370$ \kms, is similar to that
of nearby radio galaxies.  Using the correlation between stellar
velocity dispersion and black hole mass defined for nearby galaxies,
we derive estimates of the black hole masses in the range
$10^{7.9}-10^{9.2}$ \msun.  We do not find any significant difference
between the black hole masses in high-frequency-peaked and
low-frequency-peaked BL Lac objects.  Combining the velocity
dispersions with previously measured host galaxy structural
parameters, we find that the host galaxies lie on the fundamental
plane of elliptical galaxies.  This supports the conclusions of
imaging studies that the majority of BL Lac hosts are normal giant
ellipticals.

\end{abstract}

\keywords{BL Lacertae objects: general --- galaxies: active ---
  galaxies: elliptical and lenticular --- galaxies: kinematics and
  dynamics --- galaxies: nuclei}

\section{Introduction}

One of the goals of AGN research is to develop a unified framework in
which the diversity of the AGN family might be understood in terms of
variations in a few fundamental parameters such as the black hole
mass, the ratio of the accretion rate to the Eddington accretion rate,
and the orientation relative to our line of sight
\citep[e.g.,][]{lao00, bor02}.  Determining the masses of the central
black holes in different classes of AGNs is an important step toward
this goal.  An key advance in this area was the recent discovery of
the \msigma\ relation, a tight correlation between black hole mass and
stellar velocity dispersion in the host galaxy bulge \citep{fm00,
geb00a}.  The \msigma\ relation has provided an important consistency
check for black hole masses determined from reverberation mapping of
Seyfert nuclei \citep{nel00,geb00b,fer01}.  The tightness of the
correlation over a wide range of host galaxy types, both elliptical
and spiral, makes it tremendously useful as a means to determine the
black hole masses in AGNs.  While stellar-dynamical or gas-dynamical
black hole mass measurements are particularly difficult for AGNs,
stellar velocity dispersions can be measured relatively easily in some
classes of active galaxies.  Furthermore, black hole mass estimates
based on velocity dispersions are expected to be much more accurate
than estimates derived from the loose correlation between \mbh\ and
bulge luminosity \citep{mag98, kg01}.

We recently began a program to measure the stellar velocity
dispersions in the host galaxies of low-redshift BL Lac objects in
order to apply the \msigma\ relation and determine their black hole
masses.  The first measurement of a velocity dispersion in a BL Lac
object, for the TeV $\gamma$-ray source Markarian 501, indicated a
likely black hole mass of $(0.9-3.4)\times10^9$ \msun\ \citep[][Paper
I]{bhs02a}.  In this paper, we report on measurements for a sample of
11 objects and discuss the results in the context of unification
models for radio-loud AGNs.  We also compare our results with
measurements of stellar velocity dispersions for BL Lac objects
reported recently by \citet[][hereinafter FKT]{fkt02}.

\section{Observations and Reductions}

The spectra were obtained at the Hale 5m telescope at Palomar
Observatory, and at the 2.5m du Pont and 6.5m Baade (Magellan 1)
telescopes at Las Campanas Observatory.  We used the Double
Spectrograph at Palomar \citep{og82}, the ModSpec Spectrograph at the
du Pont telescope, and the Boller \& Chivens Spectrograph at Magellan.
Details of the instrumental setups are given in Table 1; Table 2
describes the exposures for each object.  Conditions were mediocre
during the Palomar nights, with typical seeing of 1\farcs5 to 2\farcs0
and thick clouds at times, while the Las Campanas nights were clear.
The gratings and grating angles were chosen to cover the most useful
spectral regions for measuring velocity dispersions: the \mgb\ region
in the blue, and the \ion{Ca}{2} near-infrared triplet in the red.
The values for the seeing listed in Table 2 were measured from
observations of template and standard stars, but these were mostly
observed during twilight and the seeing during the BL Lac observations
may have been somewhat different.  For each galaxy, the slit was
oriented close to the parallactic angle at the midpoint of the
observation, and the total exposure time was divided into individual
exposures of typically 1800 s.

Bias subtraction, flat-fielding, spectral extractions, and wavelength
calibration were done using standard routines in IRAF, and the flux
calibration and subsequent measurements were performed in IDL.
Spectra were extracted using the optimal algorithm of \citet{hor86} as
implemented in IRAF; the extraction widths are listed in Table 2.  The
strongest telluric absorption bands were removed by division by the
normalized spectrum of a nearly featureless star following methods
similar to those of \citet{wh88}.  After wavelength calibration, a
final shift to the wavelength scale was applied to each exposure based
on the wavelengths of night sky emission lines.  This step ensured
that the stellar absorption features were not artificially broadened
by summing exposures with slightly offset wavelength scales.

During each observing run, several stars were observed for use as
velocity templates.  The template stars were chosen to be primarily
giants of spectral type between G8 and K5, but we also observed some
subgiants and dwarfs and also a few F and early-G stars.  In the
following analysis, we consider only the G and K giant templates,
since they give the best match to the dominant stellar populations in
the blazar host galaxies.

In addition to the sample listed in Table 1, we also observed the
object BL Lac at Palomar, but were unable to measure its velocity
dispersion.  The stellar absorption lines were only barely detected in
a total exposure of 7200 s due to overwhelming AGN contamination.
Also, BL Lac has a Galactic extinction of $A_B = 1.42$ mag
\citep{sfd98}, and the dominant features over much of its optical
spectrum are the Galactic diffuse interstellar bands \citep{ver95}.

\section{Measurement of Velocity Dispersion}

Measurement of the velocity dispersion of a BL Lac host galaxy is
straightforward in principle, but there are some complications arising
from the dilution of the host galaxy starlight by nonthermal emission.
For the observations presented here, the detected flux within the
spectroscopic aperture is $\sim50-80\%$ nonthermal continuum emission.
As a result, a BL Lac object spectrum yields a more uncertain velocity
dispersion than a normal galaxy spectrum of the same signal-to-noise
(S/N) ratio.  Furthermore, Galactic or telluric absorption features,
if not corrected for, will have a relatively larger impact on the
velocity dispersion measurements for the BL Lac objects.

We measured the velocity dispersions of the BL Lac hosts by fitting
broadened stellar spectra to the object spectra, following a method
similar to those described by \citet{vdm94}, \citet{rix95}, and
\citet{kel00}.  The object and template spectra are first rebinned to
a wavelength scale that is linear in $\log \lambda$, and the object
spectrum is transformed to zero redshift.  The template spectrum is
broadened by convolution with a Gaussian, and adjusted to match the
galaxy spectrum by adding a linear featureless continuum and
multiplying by a low-order polynomial.  The fit includes six free
parameters: the systemic velocity of the galaxy, the width of the
Gaussian broadening function, the normalization and slope of the
featureless continuum, and the three coefficients of a quadratic
polynomial.  Multiplication by the polynomial allows the fit to
account for any reddening or other differences in overall spectral
shape between the galaxy and the template, without introducing any
high-frequency structure that might affect the dispersion measurement.
The fit is optimized using a downhill simplex algorithm \citep{nr}.
For each template, the fit is repeated several times with a range of
initial parameter values to ensure that the global minimum of \chisq\
is found.  The $1\sigma$ uncertainty for each template is determined by finding
how far the velocity dispersion must be varied from its best-fit value
(with all other parameters allowed to float freely) in order to
increase \chisq\ by unity relative to its minimum value. 

The wavelength range for the fits was selected to include the regions
most sensitive to the stellar velocity dispersion.  For each object,
we attempted to use the largest wavelength range over which a good fit
could be achieved, excluding emission lines or regions that
consistently gave poor fits with a range of template stars.  In
general, velocity dispersions measured from the Ca triplet lines are
preferred, if they can be observed, because the Ca triplet is less
sensitive to template mismatch than the blue spectral region
\citep{dre84}.  Also, the Ca triplet lines are intrinsically strong
and well isolated from neighboring lines, and can be detected even in
highly diluted spectra.  For the lowest-redshift objects (Mrk 421, Mrk
501, and 2201+044), we used the Ca triplet lines to measure $\sigma$.

At recession velocities beyond $\sim9000$ \kms, the \ion{Ca}{2}
$\lambda8662$ line is redshifted into deep telluric water vapor
absorption bands that begin at about 8925 \AA.  Reliable velocity
dispersions can still be measured using the Ca lines at 8498 and 8542
\AA\ out to $cz \approx 12000$ \kms.  For the objects having larger
redshifts, we used the region around \mgb\ to measure the velocity
dispersions.  Best results were obtained for the region $5205-5450$
\AA, which contains reasonably strong features due to Ca and Fe.  We
also included a wavelength range just blueward of \mgb\ in the fit,
typically $5030-5150$ \AA, for four objects having sufficient S/N in
this region. The wavelength region was adjusted for some objects to
exclude Galactic absorption features or bad pixels, or to include a
wider wavelength range when a good fit could be obtained.  

\begin{figure}
\plotone{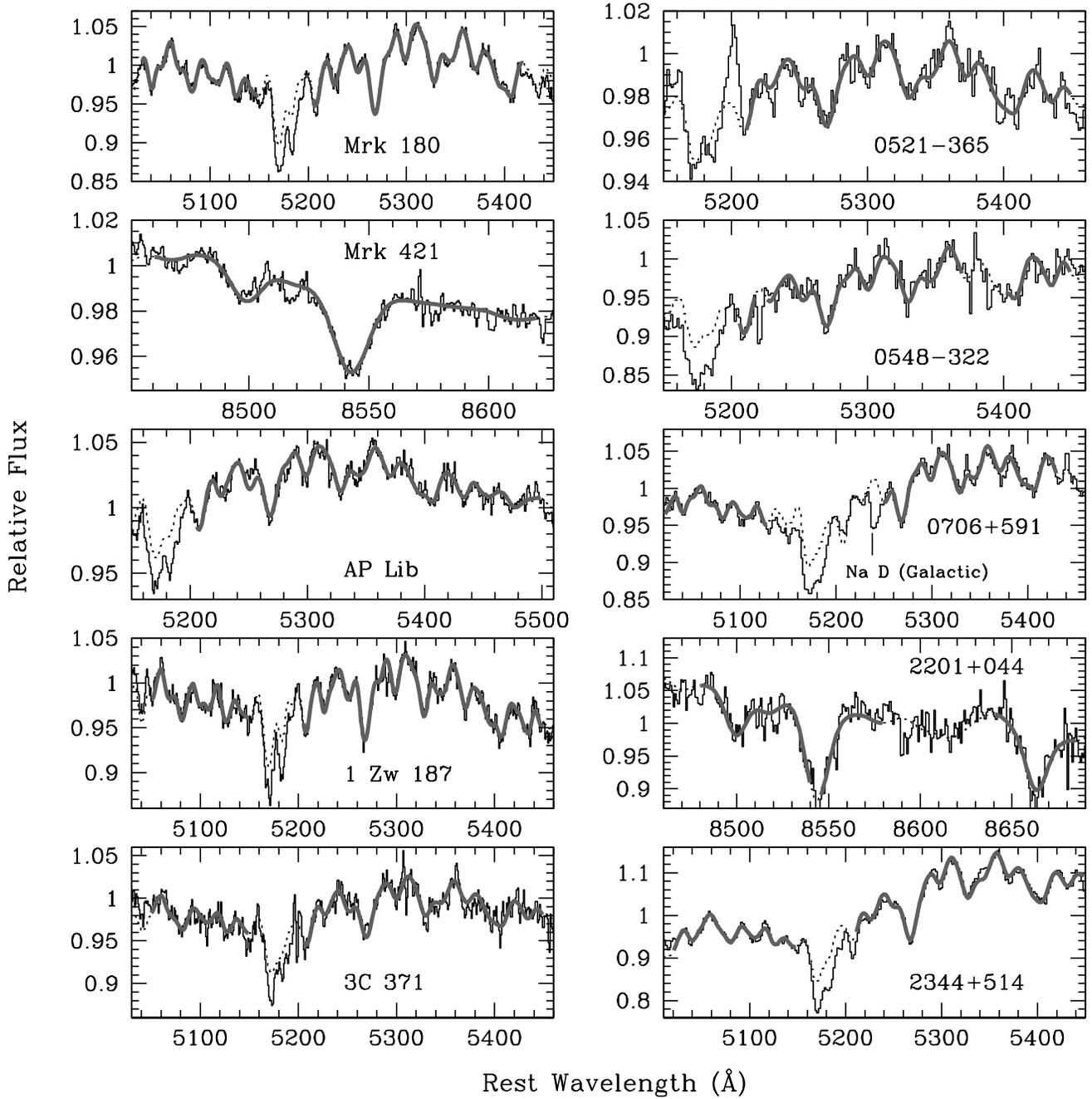}
\caption{Best fits of broadened stellar templates to the BL Lac
  spectra.  The thick curves show the broadened templates over the
  spectral region used to compute \chisq.  In regions excluded from
  the fit, the broadened template spectra are shown with a dotted
  line.  Galactic Na D absorption in the spectrum of 0706+591 is
  labelled.  Mrk 501 is shown in Paper I.
\label{fits}}
\end{figure}

It proved impossible to fit both the $5205-5450$ \AA\ region (which
contains the Fe 5270 and Fe 5335 blends) and the \mgb\ line
simultaneously.  This is a consequence of the correlation between the
[Mg/Fe] ratio and velocity dispersion in elliptical galaxies
\citep{wfg92, tra98, kun01}.  Since the Mg lines increase in strength
more steeply than do the Fe lines as a function of $\sigma$, it is
difficult to find local template stars that will provide an exact
match to the [Mg/Fe] ratio of any individual giant elliptical galaxy.
This source of template mismatch is a particular problem for direct
fitting measurements, since the calculation of \chisq\ is sensitive to
both the width and the depth of absorption lines.  Thus, we excluded
the \mgb\ line from the fits, preferring to fit the larger surrounding
region.  \citet{bhs02b} discuss the [Mg/Fe] template matching problem
in further detail, using spectra of nearby galaxies obtained during
the same observing runs.  The best fits for each object are shown in
Figure \ref{fits}.

The dominant contribution to the measurement uncertainty in most cases
is template mismatch, since the scatter in $\sigma$ measurements for
different templates is generally larger than the fitting uncertainty
for a single template.  The final velocity dispersions listed in Table
2 are the weighted mean of values measured with several templates of
type G8--K5.  To account for template matching errors, the measurement
error for each galaxy is the sum in quadrature of the fitting error
for the best-fitting template and the standard deviation of the
measurements for all templates.

There is one additional source of systematic uncertainty that
contributes significantly to the error budget.  As discussed by
\citet{bhs02b}, tests on high S/N spectra of nearby galaxies
demonstrate that our fitting results for the \mgb\ region are somewhat
sensitive to the order of the polynomial that is used to match the
continuum shape of the template to the galaxy.  For polynomial orders
of 2, 3, or 4, we find variations in the derived value of $\sigma$ at
the 5\% level.  To account for this systematic uncertainty in the
model fits, we add $0.05\sigma$ in quadrature to the uncertainties on
each measurement of $\sigma$ derived from the blue spectral region.
The Ca triplet measurements are much less sensitive to the polynomial
order, with variations at the level of $\lesssim1\%$.

Another potential source of systematic error is the variation of the
instrumental linewidth across the detector, since a given spectral
feature will fall on a different portion of the CCD for different
redshifts.  This would have the largest effect on the Palomar blue
side measurements with the 600 line grating, which was the setting
with the lowest spectral resolution, and also with the largest focus
variations across the detector.  From examination of comparison lamp
spectra, we estimate that the maximum variation in instrumental
dispersion over the region used for the measurements is $\sim15$ \kms\
in this setting.  Since the instrumental dispersion contributes in
quadrature to the total observed linewidth, these variations will have
a very small impact on the final measurements.  For the
higher-dispersion settings the impact of focus variations on the
measurements will be negligible.

\subsection{Tests of the Measurement Routine}

At Palomar, we also observed six velocity dispersion standard galaxies
from the catalog of \citet{mce95}, with $\sigma$ ranging over
$\sim100-300$ \kms, as a check on the fitting routine.  The results
are listed in Table 3.  The blue and red side measurements for each
galaxy agree within their $1\sigma$ uncertainty ranges.  We find good
agreement between our measurements and the average velocity dispersion
data compiled by McElroy for four of the six galaxies, while our
dispersions for NGC 4579 and NGC 4736 are significantly lower than the
McElroy average values.  However, in both cases our measurements for
these objects do fall in the range of previously measured values as
listed in the Hypercat database,\footnote{The Hypercat database can be
found at http://www-obs.univ-lyon1.fr/hypercat/ .} and the close
agreement between the blue and red measurements suggests that our
values of $\sigma$ do give an accurate measure of the velocity
dispersion within the observed aperture.  We believe that the
differences between our results and the mean catalog results for these
galaxies are due to individual, discrepant measurements that are
included in the catalog average velocity dispersions. A full
description of the nearby galaxy measurements is given by
\citet{bhs02b}.

Our observations have a wide range in S/N, and one possible concern is
that at low S/N the fitting routine could be biased in a systematic
way.  To test the accuracy of the results, we performed Monte Carlo
simulations to determine how well we could recover the correct
velocity dispersion as a function of S/N. Starting with the Palomar
blue side spectrum of NGC 2841, which was observed with the 1200 line
grating, we added a featureless $f_\lambda \propto \lambda^{-1}$
continuum normalized to contribute 2/3 of the total flux at 5500 \AA.
Then, Poisson noise was added to the spectrum to achieve a desired S/N
ratio.  For each value of the S/N ratio (ranging from 500 to 50 per
pixel), we created 100 realizations of the diluted, noise-added
spectrum and measured their velocity dispersions using the same
measurement routine and fitting over the wavelength ranges 5050--5150
and 5205--5450 \AA.  Ten template stars were used for each
measurement.

\begin{figure}
\plotone{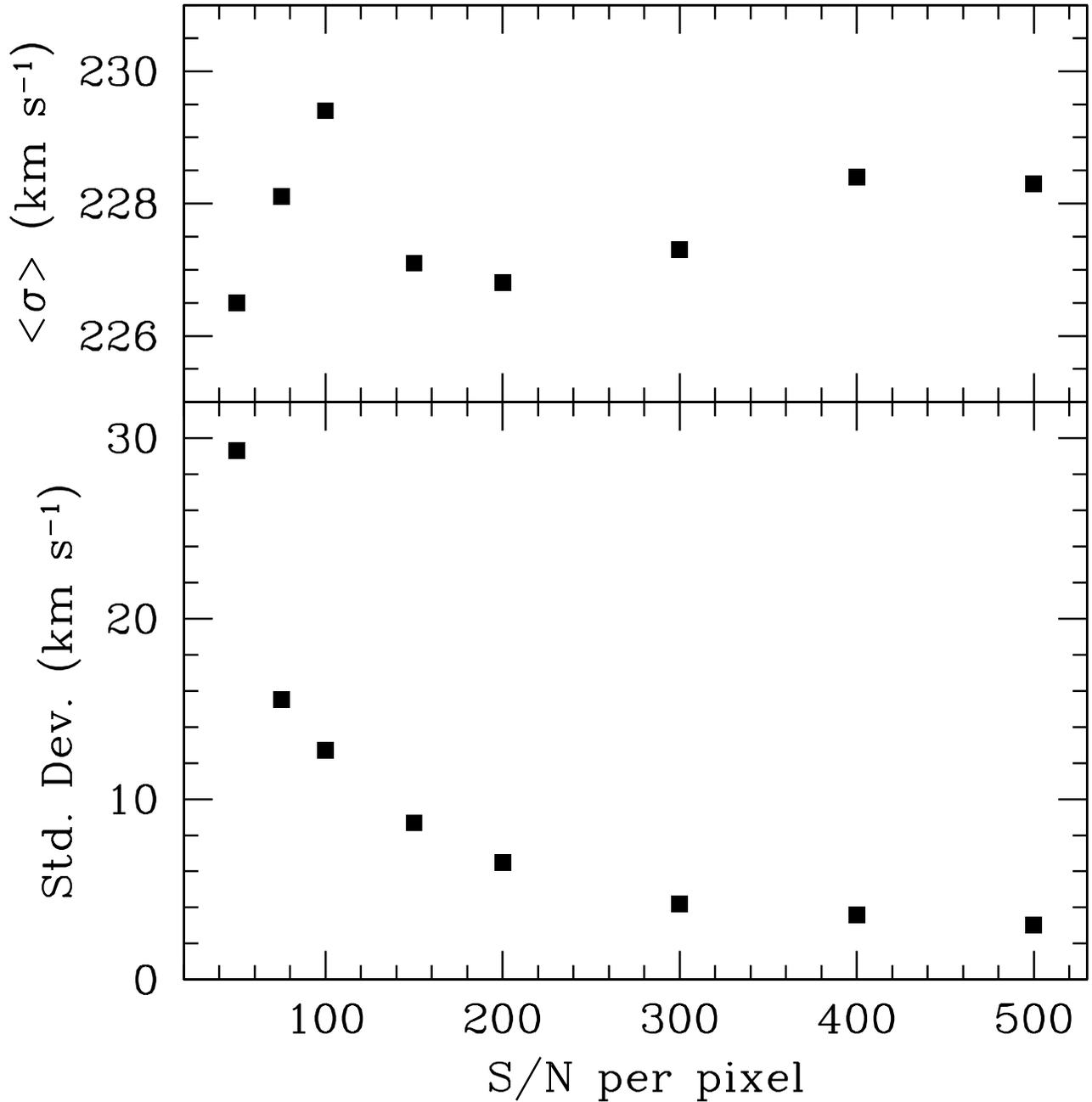}
\caption{Results of Monte Carlo simulations for measurement of the
  velocity dispersion from the diluted, noise-added spectrum of NGC
  2841.  A power-law continuum was added to contribute 2/3 of the
  total flux at 5500 \AA, and Poisson noise was added to yield a given
  S/N per pixel.  \emph{Upper panel:} The weighted average velocity
  dispersion measured from 100 realizations of the noise-added
  spectrum at each S/N value, with 10 template stars used for each
  measurement.  Without dilution or added noise, the measured velocity
  dispersion is $229 \pm 6$ \kms.  \emph{Lower panel:} The standard
  deviation of velocity dispersions measured from the 100 trials at
  each S/N value.
\label{montecarlo}}
\end{figure}

Measurement of the original spectrum gives $\sigma = 229 \pm 5$ \kms,
in good agreement with previous results from the literature
\citep{mce95}.  The results of the Monte Carlo calculations are
displayed in Figure \ref{montecarlo}.  At each S/N ratio, we computed
the weighted average and the standard deviation of the velocity
dispersion measurements for the 100 realizations of the spectrum.
Over the full range of S/N used in the calculations, the mean velocity
dispersion is within $\pm3$ \kms\ of the expected result (229 \kms),
demonstrating that there is no systematic offset as a function of S/N.
As expected, the scatter in the measurements rises at low S/N, but for
S/N > 100 per pixel the standard deviation is less than 15 \kms.  All
but one of our observations have S/N > 100 per pixel, so the expected
measurement error is not large.  The actual expected error for any
measurement is a function of the S/N, dilution, spectral resolution,
and velocity dispersion, and we do not attempt to calculate it in more
detail, but the simulations demonstrate that the template fitting
routine should work well for our data.

\subsection{Notes on Individual Objects}
\label{notes}

\emph{Mrk 421:} The Ca triplet lines are easily detected and we found
a good fit to the $\lambda8542$ \AA\ profile for $\sigma=216$ \kms.
The $\lambda8662$ line was not used in the fit as the telluric H$_2$O
absorption beyond 8900 \AA\ could not be adequately removed from the
spectrum by standard star division.  The blue side spectrum was more
strongly diluted by nonthermal light than the other galaxies in the
sample, and it proved difficult to obtain an adequate fit.  For five
template stars, the blue side data yields $\sigma = 213 \pm 47$ \kms.
Thus, we use the red side measurement as our final result.

\emph{Mrk 501:} The \ion{Ca}{2} triplet region gives $\sigma = 372 \pm
18$ \kms, where the uncertainty range includes the scatter among 24
template stars ranging between F3 and K5 and including giants,
subgiants, and dwarfs (Paper I).  In the blue spectral range, the
best-fitting template (HD 125560, a K3III star) gives $\sigma=386\pm9$
\kms\ over 5200-5600 \AA, and the standard deviation of results with
\emph{all} templates is 81 \kms.  However, if we restrict the fitting
to only G and K giants, the scatter is reduced to $\pm11$ \kms.

During the night that Mrk 501 was observed, the blue side flatfield
structure shifted on the detector along the dispersion direction,
presumably due to mechanical flexure in the spectrograph.  We observed
dome and internal flats in the afternoon.  Objects observed during the
first half of the night flattened well, but in objects observed during
the second half of the night (including Mrk 501) there were some
residual bumps in the flattened images at the $\sim10\%$ level at
$\lambda < 5000$ \AA.  Although this did not visibly affect the region
used to measure $\sigma$, and the blue and red side measurements are
in agreement, we consider the red side measurement to be more reliable
and we use it as our best estimate of $\sigma$.

A \emph{Hubble Space Telescope} (\hst) image of Mrk 501 was taken as
part of the snapshot survey of \citet{sca00} and \citet{urr00}, but it
was not included in the analysis of their \hst\ sample.  To measure
the structural parameters of the host galaxy, we downloaded the images
from the \hst\ data archive.  There are 5 separate exposures with
exposure times of 2 s, $2\times30$ s, and $2\times120$ s, all with the
F702W filter (similar to ground-based $R$).  The nucleus is saturated
in the 30 s and 120 s exposures.  We combined the 120 s exposures,
eliminating cosmic-ray hits, and used the \textsc{galfit} program
\citep{pen02} to perform a two-dimensional fit to the host galaxy
profile.  The model used in the fit was a \citet{dev48} profile
combined with an unresolved nuclear point source, convolved with a
synthetic point-spread function (PSF) generated by the Tiny Tim
program \citep{kh01}.  Saturated pixels at the nucleus were masked out
in the fits to the longer exposure.  For consistency with the
\citet{urr00} sample, we converted the F702W filter magnitude to the
Cousins $R$ band with the \textsc{synphot} package in IRAF, using a
redshifted elliptical galaxy spectrum to calculate the correction.
The total host galaxy magnitude is $m_R = 13.07$ mag, in good
agreement with the ground-based measurement of \citet{wsy96}.  The fit
gives an effective radius of $r_e = 4\farcs4$, substantially smaller
than previous ground-based measurements \citep{nil99}; the comparison
between ground-based and \hst\ measurements of \reff\ is discussed
further in \S\ref{discussion}.

\emph{0521--365 and 0548--322:} During the Magellan run, only one
K-giant stellar template was observed (HR 1629, K4III). To allow for
template mismatch error, we add 20 \kms\ in quadrature to the
uncertainty in the velocity dispersion measurement from this template.

\emph{0706+591:} At $z=0.125$, this is the most distant object in the
sample.  Galactic Na D absorption falls in the region used to
measure $\sigma$ and is excluded from the fit.  The broad H$_2$O
telluric absorption band between 5870 and 6000 \AA\ also falls in this
region.  We could not entirely exclude this band from the fit as it
covers the most useful portion of the spectrum; inclusion of part of
the telluric band in the fitting region is probably responsible
for the relatively poor fit in the neighborhood of the \mgb\ line.
The measurement yields $\sigma = 216\pm23$ \kms.  As a check on this
result, we also performed a fit over the region 4920--5150 \AA\
(excluding a 20 \AA\ region around [\ion{O}{3}] $\lambda5007$ although
this line was not clearly detected), and found $\sigma =204\pm27$
\kms.

\emph{2201+044:} The core of the \ion{Ca}{2} $\lambda8542$ absorption
line falls on a night sky emission line and the extracted spectrum has
a sharp negative ``spike'' at this wavelength.  We masked out a 5
\AA-wide region surrounding this spike in the final fit, but the
measured value of $\sigma$ changes by less than 10\% if the entire
line is included in the fit.  Over the wavelength range used for the
fit there are several extremely weak telluric absorption features with
expected equivalent widths less than 0.02 \AA.  They were not masked
out in the fit, since their amplitudes are smaller than the noise
level in the data.

\subsection{Comparison with the results of Falomo \etal}
\label{falomo}

FKT reported velocity dispersion measurements for seven BL Lac
objects, six of which are in our sample.  Comparison of the results
shows close agreement for some galaxies such as Mrk 421, but also some
surprisingly large differences.  The worst disagreement is found for Mrk 501
and 1 Zw 187.  For these two galaxies, FKT find $\sigma_c$ = 291 and
253 \kms, respectively, while we obtain $\sigma$ = 372 and 171 \kms.
FKT suggest that the different results for Mrk 501 could be the result
of telluric absorption features in our red spectrum.  However, this is
not the case because we excluded these telluric lines from the fitting
region, as described in Paper I.  Different aperture sizes cannot be the
cause of such large discrepancies either.  FKT used a 1\arcsec\ slit
and extraction widths of $3\arcsec-5\arcsec$, similar to the aperture
sizes we used, and the resulting variations in $\sigma$ should be
below the 10\% level in most cases \citep{geb00a, mf01}.

FKT used the Fourier quotient routine in IRAF which, in principle,
should yield the same results on the same data as a direct fitting
routine.  However, their measurements were carried out over a much
wider wavelength range containing different spectral features.  They
used two grating settings for their measurements: 4800--5800 \AA\ and
5700--8000 \AA\ (setups ``A'' and ``B'', respectively).  These ranges
include \hal\ and \hbeta, which are among the strongest absorption
features in the spectrum but which also appear in emission in some
objects.  They also contain other possible emission features including
[\ion{O}{3}] $\lambda5007$.  (In our sample of 11 objects,
[\ion{O}{3}] emission is clearly visible in five, and \hbeta\ in
three.)  FKT removed emission lines from their spectral by connecting
a linear continuum across the wavelength range spanned by the emission
lines.  This is a very poor approximation to the true starlight
spectrum of the galaxy, particularly for the Balmer lines, and may
adversely effect the results.  It would have been preferable to
replace the emission lines with the spectra of appropriately broadened
template stars \citep{vf93} or with spectra of galaxies having similar
velocity dispersion and line strength \citep{dal91}, or instead to
perform the fit over a wavelength range that is devoid of emission
lines.  Even when \hal\ and \hbeta\ emission are not clearly visible
in the spectra, they could still partially fill in the Balmer
absorption lines and affect the measurements.

The strongest absorption feature in the wavelength range used by FKT
is the Na D doublet at 5890, 5896 \AA, but this could have a
contribution from interstellar absorption in the host galaxies.  There
is Galactic interstellar Na D absorption in some objects as well; our
spectra reveal strong Galactic Na D absorption in Mrk 421, and weak
absorption in 0706+591.  With the Fourier quotient routine there is no
entirely satisfactory way to exclude such a feature from the
measurement.  Since there are few other strong stellar features within
a few hundred \AA\ of Na D, this spectral region is considered to be
one of the least useful portions of the optical spectrum for measuring
velocity dispersions \citep{dre84}.  The weak telluric O$_2$ and
H$_2$O absorption bands between 5800 and 6600 \AA\ might also have a
small effect on the measurements, especially in objects with a very
high degree of dilution by nonthermal light.\footnote{See Fig. 16 of
\citet{mat00} for a plot of the telluric absorption spectrum over this
wavelength range.}

Whenever possible, our measurements were performed on regions selected
to have good sensitivity to velocity dispersion without having
significant contamination from nebular emission, interstellar
absorption, or telluric absorption.  A major advantage of the direct
fitting method over the Fourier quotient method for these objects is
that it provides direct visual feedback on the quality of the template
match.  This makes it possible to mask out regions which are not fit
well by any template star, as well as telluric absorption bands and
emission lines.  We note, also, that our spectra have considerably
higher S/N than those obtained by FKT; this enables us to perform our
velocity dispersion measurements on a smaller spectral region.  In
summary, we believe that our velocity dispersion determinations are
more accurate than those of FKT.
 
\section{Black Hole Masses}
\label{bhmass}

To determine \mbh\ from the velocity dispersion, we use the recent fit
to the \msigma\ relation from \citet{tre02}, which is given by
\begin{displaymath}
\log(\mbh/\msun) = (8.13 \pm 0.06) + (4.02 \pm 0.32)
\log\left(\frac{\sigma}{200 \mathrm{~km~s\per}}\right).
\end{displaymath}
This relation was derived using a carefully culled sample of the
best-quality BH mass measurements, and using a fitting technique that
properly accounts for uncertainties both in \mbh\ and $\sigma$. 

The measurements used to calibrate the above relation were
luminosity-weighted velocity dispersions $\sigma_e$ measured in a slit
aperture of length 2\reff.  However, our values of $\sigma$ do not
exactly correspond to this aperture size, and our spectra generally do
not have sufficient sensitivity to measure $\sigma$ outside the inner
few arcseconds.  Velocity dispersions can be corrected to a standard
aperture size using the relations given by \cite{jfk95a}, but these
relations are not calibrated for long-slit apertures such as those
used by \citet{tre02}.  Another difficulty is that the corrections
depend on the effective radii, which are not known accurately for
these objects.  Rather than apply an uncertain correction to the
measured velocity dispersions, we choose to use our measured
dispersions directly to estimate \mbh.  As shown by \citet{geb00a},
the value of $\sigma$ measured from slit apertures of different length
will differ from $\sigma_e$ by less than 10\% for typical, nearby
elliptical galaxies.

The results are listed in Table 2, with uncertainties in \mbh\
determined by propagating the measurement uncertainty in $\sigma$ as
well as the uncertainties in the coefficients of the \msigma\ relation
from \citet{tre02}.  We do not include the additional possible error
due to intrinsic scatter in the \msigma\ relation, because the amount
of scatter is not yet well determined.  \citet{tre02} have determined
that this intrinsic scatter is $\leq0.3$ dex; it is hoped that
future BH mass measurements with \hst\ will provide a more definitive
result.  The masses range from $7\times10^7$ to $1.6\times10^9$ \msun,
with a median value of $1.8\times10^8$ \msun.

We emphasize that these black hole masses are indirect estimates, not
dynamical measurements of the mass. Nevertheless, there is no strong
reason to believe that the \msigma\ correlation would not apply to BL
Lac objects.  Imaging of BL Lacs with \hst\ has demonstrated that the
host galaxies are all apparently normal giant elliptical galaxies
\citep{sca00, urr00}, in contrast to the disturbed morphologies seen
in some QSO host galaxies \citep[e.g.,][]{bah97}.  Thus, the stellar
velocity dispersions in BL Lac objects should not be affected by
recent mergers or other peculiarities.  No nearby elliptical galaxies
have yet been found to deviate strongly from the \msigma\ relation, so
it would be very surprising if the BL Lac hosts did not follow the
correlation.  Furthermore, the upper end of the \msigma\ relation is
largely calibrated with FR I radio galaxies, including M87 and NGC
4261, which presumably would appear as BL Lac objects if viewed
along the jet, and there is no evidence for any offset in the \msigma\
correlation of radio galaxies in comparison with inactive ellipticals.
Still, it would be a useful consistency check to perform
stellar-dynamical measurements of \mbh\ in a few of the
lowest-redshift BL Lac objects to confirm that these assumptions are
indeed valid.

\section{Discussion}
\label{discussion}

According to unified models for radio-loud AGNs, BL Lac objects are FR
I radio galaxies viewed along the jet axis \citep[e.g.,][]{up95,
urr00}.  If radio galaxies and BL Lac objects were derived from
identical parent populations, then the distribution of black hole
masses in BL Lac objects should be the same as that in nearby FR I
radio galaxies.  However, different selection effects for BL Lacs and
radio galaxies might be expected to bias the BL Lac objects toward
less massive and less luminous host galaxies. For example,
\citet{urr00} point out that radio-loud AGNs with jets beamed in our
direction might not be identified as BL Lac objects if they occur in
very bright host galaxies so that their stellar continua are not diluted
enough by nonthermal emission to match the standard defining criteria
for classification as BL Lac objects.  Thus, the naive expectation
that BL Lac objects and radio galaxies should have identical
distributions of black hole masses or other host galaxy properties is
likely to be an oversimplification.  Such differences do not violate
the basic premise of unified models, as long as objects classified as
BL Lacs would still appear to be normal radio galaxies when viewed
from an oblique angle.

\begin{figure}
\plotone{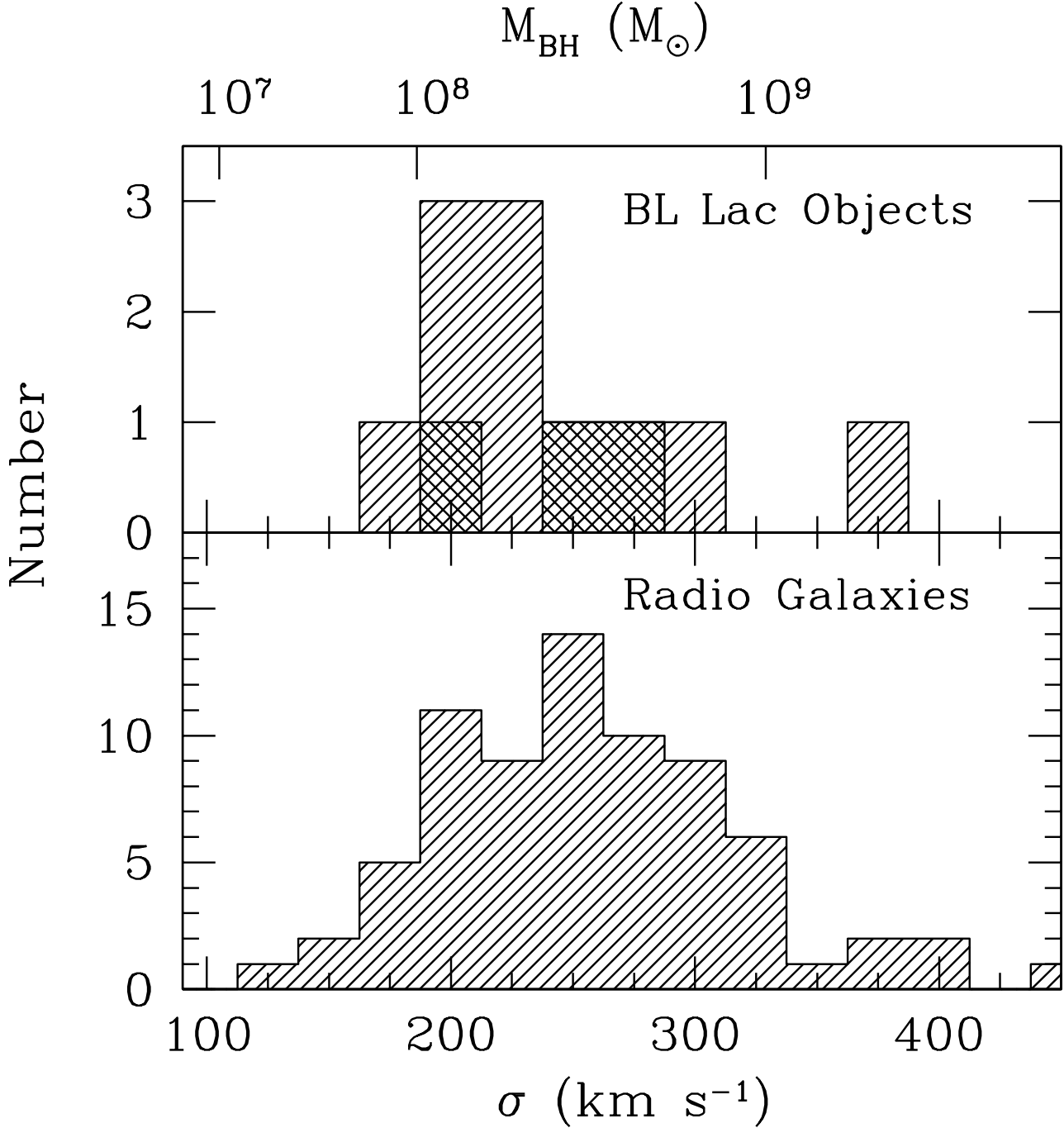}
\caption{\emph{Upper panel:} Histogram of stellar velocity dispersions
  in BL Lac objects.  Cross-hatched points indicate the
  low-frequency-peaked objects. \emph{Lower panel:} Low-redshift radio
  galaxies from the extended sample of Bettoni \etal\ (2001), which
  compiles data from several radio galaxy samples in the
  literature. The black hole mass scale at the top of the figure is
  based on the fit to the $\mbh-\sigma$ relation by \citet{tre02}.
\label{hist}}
\end{figure}

As a comparison sample of nearby radio galaxies, we use the
compilation by \citet{bet01,bet02}.  While this is not a complete sample in
any sense, it is still the largest sample of nearby radio galaxies for
which velocity dispersions, effective radii, and surface brightnesses
are available.  Figure \ref{hist} shows the distribution of velocity
dispersions for the BL Lac objects and the radio galaxy sample of
\citet{bet01}.  As demonstrated previously by FKT, the radio galaxies
and BL Lac objects occupy a similar range of values of $\sigma$, and
therefore \mbh.  The two samples have nearly the same mean velocity
dispersion, 247 \kms\ for the BL Lac objects and 256 \kms\ for the
radio galaxies.  A Kolmogorov-Smirnov test \citep{nr} on the two
distributions shows that the hypothesis of identical parent
populations for the two distributions can only be rejected at the 9\%
confidence level.  Thus, the two distributions appear to be
indistinguishable within the limited statistics, but since sample
incompleteness and selection effects are not accounted for in any way,
we caution that this comparison does not lead to any firm conclusions
regarding the nature of the two populations.

Our sample consists of three low-frequency-peaked (LBL) and eight
high-frequency-peaked (HBL) objects.  Figure \ref{hist} illustrates
that, in agreement with FKT, there is no large systematic offset
between the black hole masses in LBL and HBL objects.  The LBL and HBL
objects in our sample have <$\sigma$> = 238 and 235 \kms,
respectively.  Similarly, \hst\ images do not reveal any differences
in the host galaxy magnitudes of these blazar subtypes either
\citep{urr00}.

The \hst\ imaging survey of BL Lac objects by \citet{sca00} and
\citet{urr00} demonstrated that the host galaxies appear to be
morphologically normal ellipticals.  The morphological and dynamical
properties of elliptical galaxies are linked via the fundamental plane (FP)
relations \citep{dd87, dre87}. Therefore, by combining the velocity
dispersions with magnitudes and effective radii, we can test whether
the BL Lac host galaxies fall on the same FP as normal
ellipticals.  For consistency with \citet{urr00} and \citet{bet01}, we
use $H_0 = 50$ \kms\ Mpc\per.

One source of uncertainty in such a comparison comes from the fact
that the effective radii measured from ground-based images and from
the \hst\ snapshot survey of \citet{sca00} and \citet{urr00} are in
severe disagreement for some of the galaxies in our sample.  Table
\ref{radii} lists the effective radii from the \hst\ survey, and from
ground-based imaging from various sources in the literature.  The
ground-based measurements of \reff\ are nearly always larger than the
\hst\ measurements.  In the worst case (3C 371), the disagreement is a
factor of 5.

The main advantage of the \hst\ images is the narrow PSF core which
makes it possible to determine the point-source flux accurately, but
the \hst\ measurements are hindered by the small field of view of the
WFPC2 Planetary Camera (PC) detector (35\arcsec\ square), and brief
exposures of 200 to 1300 s.  High-quality ground based imaging
\citep[e.g.,][]{nil99} demonstrates that the host galaxies of the
nearest objects are indeed large enough to completely fill the PC
chip.  As a result, the sky background level determined from the outer
portions of the detector will be contaminated by galaxy light, and
measurements of the galaxy radial profile after subtraction of an
artificially high sky background would lead to an underestimate of
\reff.  This systematic effect could be largely responsible for the
differences between the ground-based and \hst\ measurements.  The
total host galaxy magnitudes are in better agreement, with differences
of typically less than a few tenths of a magnitude.  The problem of
determining the host galaxy parameters for BL Lac objects from shallow
\hst\ images is somewhat analogous to the difficulty of detecting QSO
host galaxies in short WFPC2 exposures, as discussed by \citet{mr95}.
Nevertheless, it is clear that the ground-based measurements of \reff\
are quite uncertain as well. The case of Mrk 501 illustrates the
difficulty of measuring \reff\ from the ground.  Even though this is
one of the nearest and best-resolved BL Lac objects, ground-based
measurements of \reff\ range from 9\arcsec\ to 20\arcsec.

\begin{figure}
\plotone{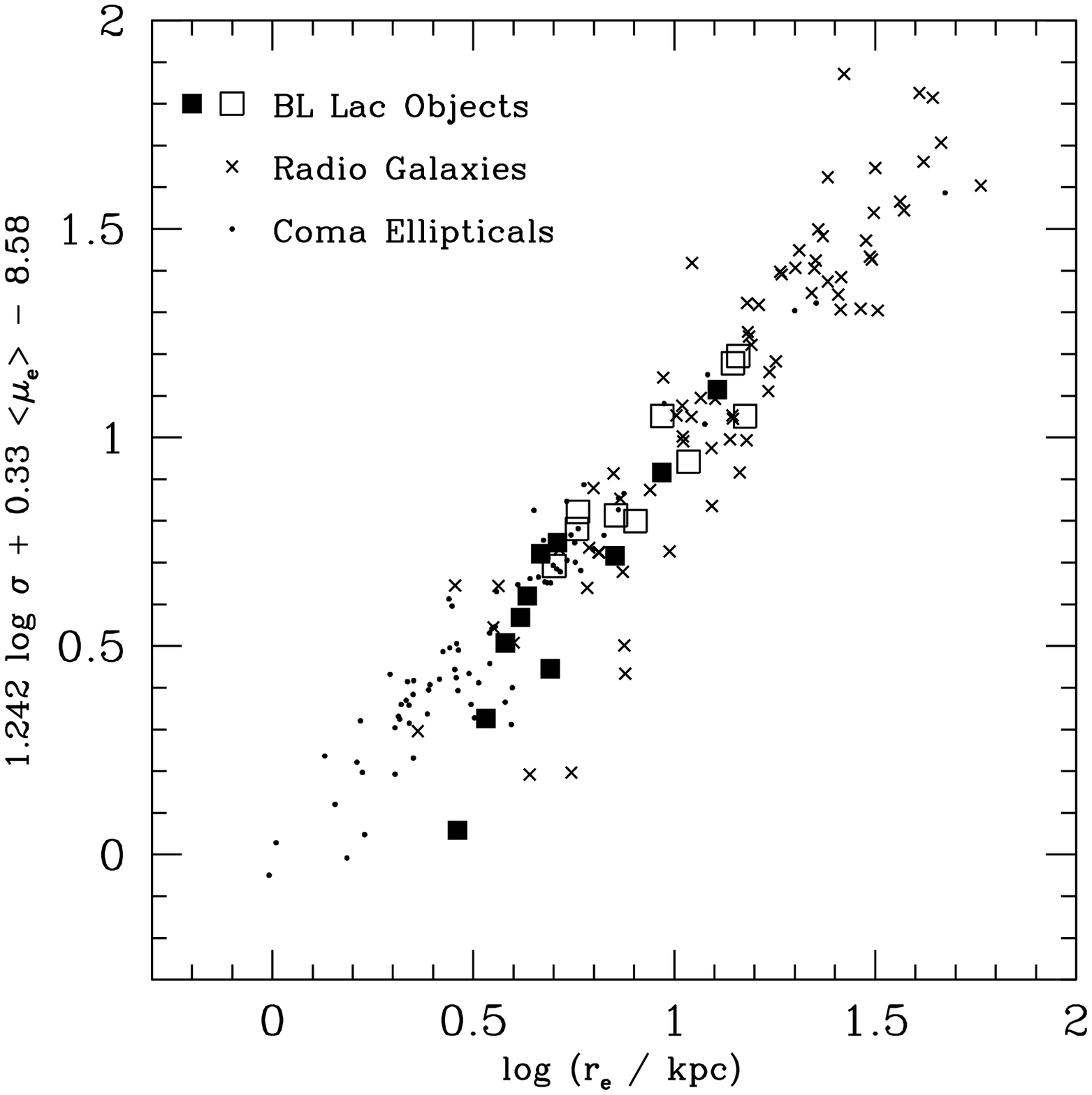}
\caption{The fundamental plane using the parameterization given by
  \citet{bet01}.  \emph{Filled squares:} BL Lac objects, using
  effective radii and magnitudes from the \hst\ survey of
  \citet{urr00}.  \emph{Open squares:} BL Lac objects using effective
  radii and magnitudes from ground-based data.  No ground-based
  measurments are available for object 0706+591.  \emph{Crosses:}
  Radio galaxies from \citet{bet01}. \emph{Small circles:} Coma
  cluster ellipticals from \citet{jfk95a,jfk95b}.
\label{fplane}}
\end{figure}

Figure \ref{fplane} plots the FP for the BL Lac objects
using the parameterization given by \citet{bet01} for low-redshift
radio galaxies.  Given the uncertainty in the measurements of \reff,
and the consequent uncertainty in \mue\ (the average $R$-band
surface brightness enclosed within \reff), we plot the BL Lac objects
using both the ground-based and \hst\ measurements.  The host galaxy
magnitudes have been corrected for Galactic extinction and
$k$-corrected as described by \citet{urr00}.  The ground-based data
included in the plot are from \citet{amc91}, \citet{wsy96}, and
\citet{nil99}.  For comparison, the radio galaxy sample of
\citet{bet01} and the Coma cluster ellipticals observed by
\citet{jfk95a, jfk95b} are also shown.  

The overall agreement is very reasonable, considering the relatively
large uncertainties in the measurements of $\sigma$ and \reff\ for the
BL Lac objects.  Changes in $r_e$ at fixed total magnitude result in
displacement along a vector that is nearly parallel to the FP
sequence; as a result both the ground-based and \hst-based
measurements appear consistent with the FP.  In either case, most of
the BL Lac objects fall within the scatter of the FP. The largest
outlier is 3C 371 when plotted with the effective radius of 2.9 kpc
from \citet{urr00}; it deviates from the FP sequence by $\sim0.3$ dex.
However, ground-based measurements indicate an effective radius that
is larger by a factor of five, and with the ground-based value of
\reff\ this galaxy falls nicely on the FP.  Thus, despite the
uncertainty in the effective radii, the host galaxies do appear to be
normal elliptical galaxies.

\begin{figure}
\begin{center}
\plotone{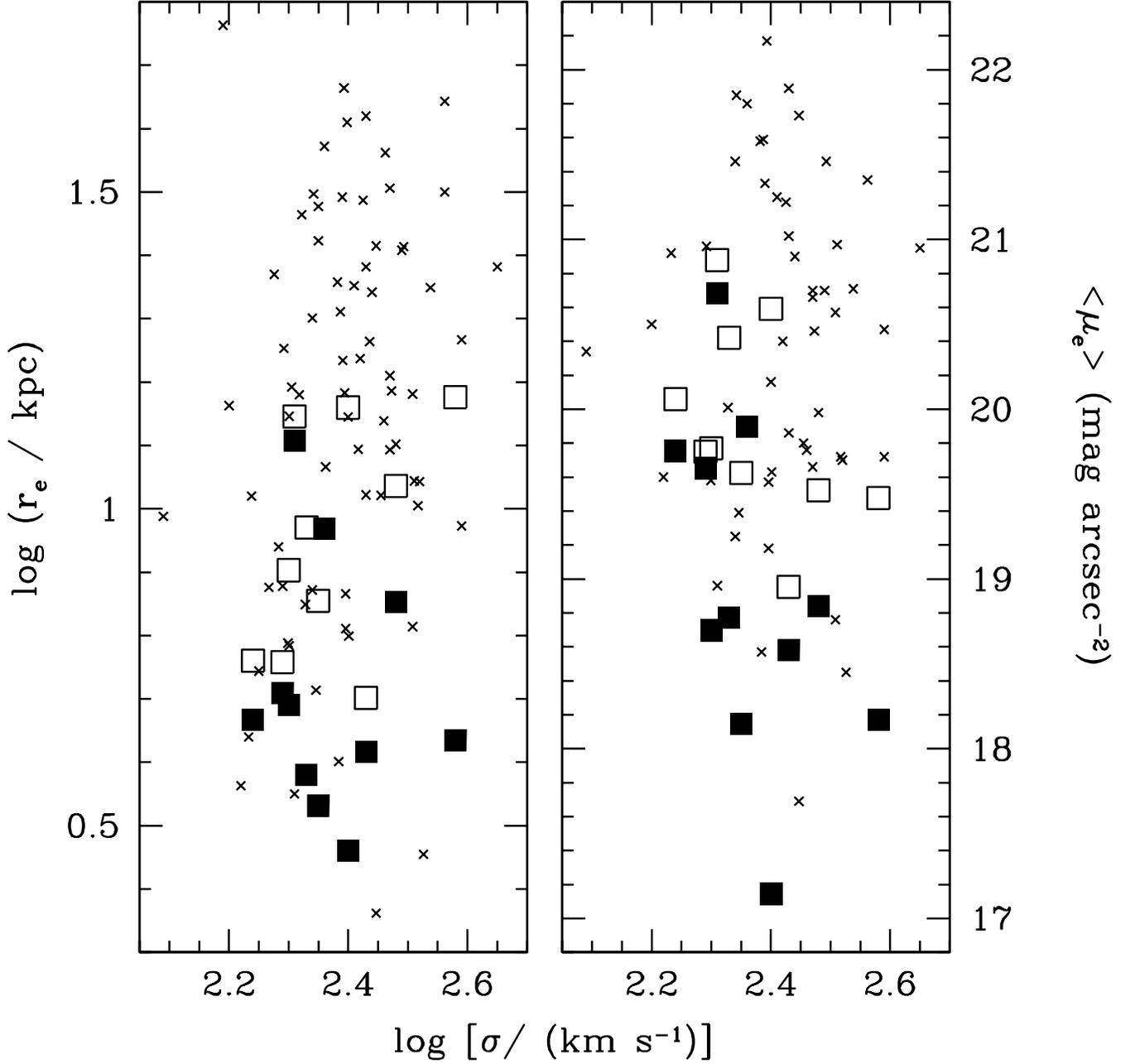}
\end{center}
\caption{Projections of the fundamental plane onto the $\log \sigma$
  vs. $\log r_e$ and \mue\ planes.  \emph{Filled
  squares:} BL Lac objects, using effective radii and magnitudes from
  the \hst\ survey of \citet{urr00}.  \emph{Open squares:} BL Lac
  objects using effective radii and magnitudes from ground-based data.
  \emph{Crosses:} Radio galaxies from \citet{bet01}.
\label{fproj}}
\end{figure}

The contrast between the ground-based and \hst-based measurements of
\reff\ and \mue\ can be seen more clearly by plotting projections
of the FP onto the \reff\ and \mue\ axes, as shown
in Figure \ref{fproj}.  With the \hst\ measurements, the BL Lac
objects have effective radii comparable to the smallest radio
galaxies, and most of the BL Lac hosts have \mue\ brighter than 19
mag arcsec\persq, in a region where very few radio galaxy hosts lie.
The ground-based measurements place the BL Lac hosts in a region that
overlaps much more closely with the radio galaxies, better matching
expectations from unified models.  However, the BL Lac objects are
still displaced toward smaller radii than the largest radio galaxies;
this is also apparent in Figure \ref{fplane}, where the BL Lac objects
do not extend as far up the FP sequence as the radio
galaxies.  Similarly, \citet{urr00} noted that no BL Lac hosts are
found to be as luminous as the most luminous hosts of FR I radio
galaxies.  This offset could be a consequence of selection effects, or
might indicate an environmental effect disfavoring the formation of
blazar jets in rich clusters \citep{urr00}.

The results presented here, and by FKT, strongly disagree with several
recent estimates of \mbh\ in blazars based on other methods.  Using
variability timescales to estimate the size of the emission region,
\citet{xie98} and \citet{fxb99} derive estimates of the black hole
masses for several blazars that are in the range $\sim10^6-10^8$
\msun, much smaller than the masses obtained from the \msigma\
relation.  \citet{wxw02} have proposed a method to derive \mbh\ from
the peak luminosity and frequency of the blazar emission.  They
conclude that HBL and LBL blazars have a bimodal distribution of \mbh,
with masses of $10^5-10^8$ \msun\ for HBL and $10^9-10^{11}$ \msun\
for LBL objects.  The stellar velocity dispersions imply black hole
masses that are are completely inconsistent with this conclusion.
Models that interpret periodicities in blazar light curves in terms of
orbital motion of black hole binary systems \citep{rm00, din02} also
give much smaller masses than those derived from the \msigma\
relation.

\citet{xie02} propose a new method to estimate black hole masses in BL
Lac objects, based on the assumption that the observed variability
timescale is determined by the orbital timescale near the innermost
stable orbit around the black hole.  They find masses ranging from
$10^{7.4}$ to $10^{9.2}$ for 13 objects, in the same range as the
masses we derive.  Xie \etal\ show that their \mbh\ estimates are
strongly correlated with the beamed nonthermal luminosity but not with
the host galaxy luminosity.  However, the opposite would be expected
if BL Lac host galaxies follow the $\mbh-L_{\mathrm{bulge}}$
correlation of normal galaxies.  Furthermore, \citet{urr00} show that
there is no correlation between the nuclear and host galaxy
luminosities for BL Lac objects, implying that the nonthermal
luminosity should not be tightly correlated with \mbh\ if \mbh\ scales
with host galaxy luminosity.  Only two objects are in common between
our sample and Xie \etal, so it is not possible to perform a detailed
comparison of the mass estimates, but we note that their upper limit
to the black hole mass in 3C 371 is an order of magnitude lower than
the mass we derive from the \msigma\ correlation.

\citet{wlz02} and \citet{wu02} have recently published estimates of
the BH masses in BL Lac objects by using the \hst\ imaging survey
results to obtain approximate velocity dispersions via the FP
relation, and then applying the \msigma\ relation to estimate \mbh.
The advantage of this method is that it can be easily applied to a
large number of objects, as there are over 60 BL Lac objects with
morphological parameters measured from the \hst\ survey.  However,
there are some potential disadvantages as well, since the $\sigma$
estimates are limited by the accuracy of the host galaxy
decompositions and by the intrinsic scatter in the FP.  In both cases,
the FP-derived estimates of $\sigma$ are systematically larger than
our measured values.  For objects in common with our sample, the Wu
\etal\ estimates of $\sigma$ are higher than our values by 30\% on
average, and the Woo \& Urry estimates are higher by 74\%.  The most
severe disagreement is for 3C 371; for this object Wu \etal\ predict
$\sigma = 524$ \kms\ and Woo \& Urry predict 618 \kms, in comparison
with our measured value of $249 \pm 25$ \kms.  This illustrates the
need for accurate values of \reff\ and \mue\ when applying this
method.

Determining the Eddington ratio $L/\ledd$ in BL Lac objects can give
important clues to the physical structure of their accretion flows.
It would be interesting to search for systematic differences between
LBL and HBL objects; this could yield new constraints for models of
the spectral energy distributions \citep[e.g.,][]{ghi98, bd02}.
Unfortunately, it is particularly difficult to measure the bolometric
luminosity of the unbeamed AGN core in BL Lac objects that are
dominated by the beamed nonthermal emission from the jet.  One
alternative is to use the emission lines as a measure of the ionizing
luminosity, because the emission-line flux is the only unbeamed
component of the AGN that can be directly observed.  This method is
not ideal since there is no single emission line that can be observed
in all objects; in practice the only viable alternative is to use the
lines that are detected and perform a rough conversion either to a
total broad-line region luminosity \citep{cpg97} or to a single line
such as \hbeta\ \citep{cao02}.  For our sample, broad-line fluxes are
available only for a few objects, while a few others have [\ion{O}{3}]
emission detected.  Given this small number of objects and the large
uncertainties involved in comparing the luminosities of different
emission lines, we are unable to perform any meaningful comparison of
the Eddington ratios of the HBL and LBL objects in our sample.  This
would be a useful objective for future work if stellar velocity
dispersions and emission-line luminosities can be measured for larger
samples of BL Lac objects.

\section{Conclusions}

Our conclusions are summarized as follows:

1. Stellar velocity dispersions in low-redshift BL Lac objects are
  in the range $\sim170-370$ \kms.  Using the \citet{tre02} fit to the
  \msigma\ relation, the corresponding black hole masses are $\mbh
  \approx 10^{7.9}$ to $10^{9.2}$ \msun.

2. Measurement of velocity dispersions in BL Lac objects requires
  particular care in selecting spectral regions that are sensitive to
  $\sigma$ but are not severely affected by interstellar absorption
  lines, telluric absorption bands, or emission lines.  Due to the
  dilution by nonthermal emission, high S/N (typically $\gtrsim100$
  per pixel in the extracted spectrum) is required for accurate
  measurements with a direct template-fitting method.

3. The distribution of velocity dispersions in BL Lac objects
  appears superficially similar to that of the radio galaxy sample of
  \citet{bet01}, indicating a similar distribution of black hole
  masses for the two samples.  However, neither sample is
  statistically well defined or complete in any sense and we caution
  against drawing conclusions about radio-loud AGNs in general from
  this limited comparison.

4. There does not appear to be any systematic difference between
  the black hole masses in HBL and LBL objects. 

5. The host galaxies of BL Lac objects lie on the fundamental plane of
  nearby elliptical galaxies.  We do not find any BL Lac objects near
  the top of the fundamental plane sequence in the region occupied by
  the most luminous hosts of FR I radio galaxies; this may be a
  consequence of selection effects in the identification of BL Lac
  objects.

6. For nearby, well-resolved BL Lac objects, the host galaxy effective
  radii measured in the \hst\ WFPC2 snapshot survey of \citet{sca00}
  and \citet{urr00} are systematically smaller than effective radii
  measured from deeper ground-based images.  We suggest that the
  measurements of \reff\ from the \hst\ snapshot survey images may be
  biased toward underestimates of the true radii, because they were
  derived from shallow exposures that did not detect the faint outer
  envelopes of the host galaxies that are seen in deep ground-based
  images.

\acknowledgments

We thank Swara Ravindranath for assistance with the observing runs at
Las Campanas and with the \hst\ imaging analysis, Tom Matheson for
providing some of the spectroscopic reduction software, and Meg Urry
for helpful discussions.  Research by A.J.B. is supported by NASA
through Hubble Fellowship grant \#HST-HF-01134.01-A awarded by the
Space Telescope Science Institute, which is operated by the
Association of Universities for Research in Astronomy, Inc., for NASA,
under contract NAS 5-26555.  Research by W.L.W.S. is supported by NSF
grant AST-9900733.  This research has made use of the NASA/IPAC
Extragalactic Database (NED) which is operated by the Jet Propulsion
Laboratory, California Institute of Technology, under contract with
the National Aeronautics and Space Administration.  This research was
partly based on observations made with the NASA/ESA \emph{Hubble Space
Telescope}, obtained from the data archive at the Space Telescope
Science Institute, which is operated by the Association of
Universities for Research in Astronomy, Inc., under NASA contract NAS
5-26555.

\clearpage

\begin{deluxetable}{lccccccc}
\label{observingruns}
\tablewidth{7.5in} 
\tablecaption{Observing Runs} 
\tablehead{\colhead{Observing} & \colhead{UT Date} &
\colhead{Telescope} & \colhead{Slit} & \colhead{Grating} &
\colhead{$\lambda$} & \colhead{Plate Scale} &
\colhead{$\sigma$(instrumental)} \\ \colhead{Run} & &  & \colhead{(\arcsec)} &
\colhead{(grooves mm\per)} & \colhead{(\AA)} & \colhead{(\AA\
pixel\per)} & \colhead{(\kms)\tablenotemark{a}} }

\startdata 

1 & 2001 Jun 24 & P200 & 2 & Blue: 600 & $4210-5950$ & 1.72 & 115 \\
 & & & & Red: 1200 & $8400-9070$ & 0.63 & 25 \\
2 & 2001 Nov 9 & du Pont & 1 & 832 & $8240-9800$ & 0.72 & 25 \\
3 & 2002 Jan 22 & P200 & 2 & Blue: 600 & $4600-6350$ & 1.72 & 115 \\
 & & & & Red: 1200 & $8400-9070$ & 0.63 & 25 \\
4 & 2002 Jan 24-26 & P200 & 2 & Blue: 1200 & $4940-5820$ & 0.88 & 60 \\
 & & & & Red: 1200 & $8400-9070$ & 0.63 & 25 \\
5 & 2002 Feb 19-20 & Baade & 1 & 600 & $3600-6700$ & 1.56 & 91 \\
\enddata

\tablenotetext{a}{Instrumental dispersion for a source uniformly
  filling the slit, measured from the widths of comparison lamp lines
  at 5500 \AA\ or 8600 \AA.}
\end{deluxetable}

\begin{center}
\begin{deluxetable}{lccccccccc}
\label{journal}
\tablewidth{7in} \tablecaption{Observations and Results}
\tablehead{\colhead{Object} & \colhead{Type} & \colhead{$z$} &
\colhead{Run} & \colhead{Exposure} & \colhead{Extraction} &
\colhead{Seeing} & \colhead{S/N} & \colhead{$\sigma$} &
\colhead{$\log(\mbh/\msun)$} \\ & & \colhead{} & & \colhead{(s)}
& \colhead{(\arcsec)} & \colhead{(\arcsec)} & & \colhead{(\kms)}}

\startdata
Mrk 180   & H & 0.045 & 4 & 5400  & 5.0 & 1.5--2.0 & 130 & $209 \pm 11$    & $8.20 \pm 0.11$ \\
Mrk 421   & H & 0.030  & 3 & 10800 & 5.0 & 2       & 310 & $219 \pm 11$    & $8.28 \pm 0.11$ \\
Mrk 501   & H & 0.034 & 1 & 12600 & 3.7 & 1.5      & 305 & $372 \pm 18$    & $9.21 \pm 0.13$ \\
AP Lib    & H & 0.049 & 4 & 4850  & 5.0 & 2        & 160 & $196 \pm 21$    & $8.09 \pm 0.20$ \\
1 Zw 187  & H & 0.055 & 4 & 5400  & 5.0 & 1.5--2.0 & 100 & $171 \pm 12$    & $7.86 \pm 0.14$ \\
3C 371    & L & 0.051 & 4 & 3600  & 5.0 & 3--5     & 110 & $249 \pm 25$    & $8.51 \pm 0.19$ \\
0521--365 & L & 0.055 & 5 & 1800  & 2.5 & 0.7--1.0 & 165 & $269 \pm 29$    & $8.65 \pm 0.20$ \\
0548--322 & H & 0.069 & 5 & 3600  & 2.5 & 0.7--1.0 & 130 & $202 \pm 24$    & $8.15 \pm 0.22$ \\
0706+591  & H & 0.125 & 3 & 10800 & 5.0 & 2        & 160 & $216 \pm 23$    & $8.26 \pm 0.20$ \\
2201+044  & L & 0.027 & 2 &  7200 & 4.2 & 1.0--1.5 &  50 & $197 \pm 8\phn$ & $8.10 \pm 0.09$ \\
2344+514  & H & 0.044 & 3 &  5400 & 5.0 & 2        & 170 & $294 \pm 24$    & $8.80 \pm 0.16$ \\
\enddata

\tablecomments{Blazar type is H = high-frequency peaked, or L =
  low-frequency peaked.  Classifications are from the catalog of
  Scarpa \etal\ 2000, and redshifts are from NED.  S/N is
  the average signal-to-noise ratio per pixel in the extracted
  spectrum, over the region used to measure the velocity
  dispersion. For Mrk 501, the S/N value is listed for the red side
  exposure from Paper I. Black hole masses are calculated using the
  fit to the \msigma\ relation by \citet{tre02}.}

\end{deluxetable}
\end{center}

\begin{center}
\begin{deluxetable}{lccc}
\tablewidth{4in}
\tablecaption{Nearby Galaxies}
\label{results}
\tablehead{\colhead{Galaxy} & \multicolumn{3}{c}{$\sigma$ (\kms)} \\ 
\colhead{} & \colhead{Red} & \colhead{Blue} &
\colhead{McElroy (1995)}
}
\startdata
NGC 2841 & $222 \pm 4$ & $229 \pm 5$ & $232 \pm 18$ \\
NGC 4278 & $261 \pm 8$ & $\phn251 \pm 13$ & $250 \pm 15$ \\
NGC 4321 & $\phn92 \pm 4$ & $\phn101 \pm 12$ & $\phn94 \pm 12$ \\
NGC 4374 & $308 \pm 7$ & $302 \pm 7$ & $296 \pm 13$ \\
NGC 4579 & $165 \pm 4$ & $166 \pm 8$ &  $189 \pm 4\phn$ \\
NGC 4736 & $112 \pm 3$ & $109 \pm 4$ & $136 \pm 12$ \\
\enddata

\tablecomments{Data are from the June 2001 and January 2002 Palomar
  runs.  Comparison results from the literature are taken from the
  compilation by \citet{mce95}.  The uncertainty ranges quoted by
  McElroy represent the scatter among measurements taken from the
  literature for each galaxy. }

\end{deluxetable}
\end{center}

\begin{center}
\begin{deluxetable}{lcl}
\tablewidth{5.5in}
\tablecaption{Effective Radii of BL Lac Object Host Galaxies}
\label{results}
\tablehead{\colhead{Galaxy} & \colhead{$r_e$(\hst)} &
  \colhead{$r_e$(ground) and Reference}         \\ 
  \colhead{} & \colhead{(arcsec)} & \colhead{(arcsec)}
}
\startdata
Mrk 180   & $3.10 \pm 0.02$ & 7.1 (A91) \\
Mrk 421   & $3.95 \pm 0.05$ & 11 (N99) \\
Mrk 501   & $4.44 \pm 0.03$ & 9.3 (A91), 20.0 (S93), 9.0 (W96), 13 (N99), 17.2
  (P02), 15 (FKT) \\
AP Lib    & $3.70 \pm 0.10$ & 5.7 (A91), 8.7 (S93), 6.7 (P02) \\
1 Zw 187  & $3.15 \pm 0.05$ & 3.9 (W96) \\
3C 371    & $2.10 \pm 0.10$ & 10.5 (W96), 14.9 (S93), 10.8 (P02) \\
0521--365 & $2.80 \pm 0.07$ & 3.4 (W96) \\
0548--322 & $7.05 \pm 0.15$ & 7.7 (W96) \\
0706+591  & $3.05 \pm 0.07$ & \nodata \\
2201+044  & $6.78 \pm 0.08$ & 6.4 (F96), 7.6 (W96) \\
2344+514  & $5.93 \pm 0.02$ & 8.5 (N99), 5.8 (F99) \\
\enddata

\tablecomments{\hst\ measurements are from \citet{urr00}, except for
  Mrk 501 which is described in \S\ref{notes}.  Ground-based
  measurements are taken from the following sources.  A91:
  \citet{amc91}. S93: \citet{sfk93}. W96: \citet{wsy96}. F99:
  \citet{fk99}. N99: \citet{nil99}.  FKT: \citet{fkt02}.  P02:
  \citet{pur02}.
\label{radii}}
\end{deluxetable}
\end{center}

\end{document}